\documentclass[a4paper,11pt]{article}
\usepackage[T1]{fontenc}
\usepackage{lmodern}
\usepackage[margin=27mm]{geometry}
\usepackage{amsmath,amssymb,amsthm,mathtools}
\usepackage{booktabs,tabularx,array}
\usepackage{microtype}
\usepackage[hidelinks]{hyperref}
\allowdisplaybreaks[1]
\newcommand\NMQC{\ensuremath{\mathsf{NMQC}_{\oplus}}}
\newcommand\AC{\ensuremath{\mathsf{AC}^0}}
\newcommand\ACC{\ensuremath{\mathsf{ACC}^0}}
\newcommand\ACm{\ensuremath{\mathsf{AC}^0[m]}}
\newcommand\AND{\ensuremath{\mathsf{AND}}}
\newcommand\NOR{\ensuremath{\mathsf{NOR}}}
\newcommand\Mod[1]{\ensuremath{\mathsf{Mod}_{#1}}}
\newcommand\term[1]{\emph{#1}}
\newcommand\pfs{\mathsf{pfs}_{\epsilon}}
\newcommand\tdeg{\mathsf{deg}_{\mathbb{T},\epsilon}}

\DeclareMathOperator{\fdeg}{fdeg}
\newcommand\ip[2]{\left\langle #1,#2\right\rangle}
\newcommand\fnorm[1]{\|\widehat{#1}\|_1}
\newcommand\suppsize[1]{\operatorname{sparsity}(\widehat{#1})}
\newcommand\E{\mathbb{E}}
\newcommand\sgn{\operatorname{sgn}}
\newcommand\ee{\mathrm{e}}

\theoremstyle{plain}
\newtheorem{theorem}{Theorem}
\newtheorem*{fourierrestatement}{Theorem~\ref{thm:f1}}
\newtheorem*{circuitrestatement}{Theorem~\ref{thm:intro}}
\newtheorem{proposition}[theorem]{Proposition}
\newtheorem{lemma}[theorem]{Lemma}
\newtheorem{corollary}[theorem]{Corollary}

\newtheorem{fact}[theorem]{Fact}
\theoremstyle{definition}
\newtheorem{definition}[theorem]{Definition}
\theoremstyle{remark}
\newtheorem*{remark}{Remark}

\title{Low-Degree Torus Approximations of the AND Function and \ACC\ Circuits}
\author{Shuhei Osaki \and Ryuhei Mori}
\date{}
\hypersetup{
  pdftitle={Low-Degree Torus Approximations of the AND Function and ACC0 Circuits},
  pdfauthor={Shuhei Osaki and Ryuhei Mori},
  pdfsubject={Degree bounds for torus polynomial approximation},
  pdfkeywords={torus polynomials, Fourier analysis, AND function, ACC0, circuit complexity}
}

\begin{document}
\maketitle

\begin{abstract}
Torus polynomials, real polynomials evaluated modulo one, were proposed
by Bhrushundi, Hosseini, Lovett, and Rao as a route to $\ACC$ lower
bounds.
We prove degree upper bounds for torus polynomials approximating a
function on every input.
First, for every $\epsilon\in(0,1/4)$, $\AND_n/2$ has an
$\epsilon$-approximating torus polynomial of degree
$O(\log(n/\epsilon))$, improving the previous $O(\log^2(n/\epsilon))$.
This is optimal up to constant factors for inverse-polynomial error,
by a lower bound of Krishan and Vishwanathan, and refutes two
superlogarithmic-degree conjectures of Krishan and Sarma.
Second, for every Boolean function $g$ computed by a depth-$d$,
size-$s$ $\ACm$ circuit, $g/2$ has an $\epsilon$-approximating torus
polynomial of degree at most
$C_m^d(\log s)^{\max\{1,\mu-1\}d}\log(1/\epsilon)$, where $\mu$ is the
number of distinct prime factors of $m$ and $C_m$ depends only on $m$.
This improves the bound
$(\log s)^{O_m(d^2)}(\log(1/\epsilon))^{O_m(d)}$ of Bhrushundi,
Hosseini, Lovett, and Rao, making the exponent of $\log s$ linear in
the depth and the dependence on $\log(1/\epsilon)$ linear.
Both proofs sample terms of a Fourier expansion and approximate them
by low-degree torus polynomials.
A torus approximation of degree $D$ can be written as a linear
combination of at most $(n+1)^D$ parities, called a periodic Fourier
representation, which is known to give a non-adaptive
measurement-based quantum computation with linear side-processing
using one qubit per parity.
Hence every $\ACC$ function can be computed with quasipolynomially
many qubits and inverse-polynomial error probability on every input.
Conversely, we prove that a periodic Fourier representation with $r$
parities and error probability $\epsilon$ gives a torus approximation
of degree $O(\log(r/\epsilon))$, so the two notions are equivalent.
\end{abstract}

\section{Introduction}
Low-degree polynomial approximations provide a way to prove lower
bounds for Boolean circuits. First, one shows that every small circuit
in a given class has a low-degree approximation. A degree lower bound
for a function, using the same notion of approximation, then gives a
lower bound on the circuit size. See
Beigel~\cite{beigel_polynomial_1993} and
Williams~\cite{williams:LIPIcs.FSTTCS.2014.47}.

For $\AC$, H{\aa}stad used random restrictions and his switching lemma
to prove that depth-$d$ circuits computing parity exactly require
size $\exp(\Omega_d(n^{1/(d-1)}))$ for every fixed
$d\ge2$~\cite{hastad_1986}.
Linial, Mansour, and Nisan later used the switching lemma to prove
concentration of the Fourier spectrum on low-degree terms, which gives
approximation by real polynomials in $L_2$~\cite{10.1145/174130.174138}.
Refinements of this approximation bound by H{\aa}stad and Tal imply
that depth-$d$ circuits agreeing with parity on at least two thirds
of all inputs require size $\exp(\Omega_d(n^{1/(d-1)}))$
\cite{HASTAD2001498,tal:LIPIcs.CCC.2017.15}.
For $\mathsf{AC}^0[p]$,
where $p$ is prime, Razborov and Smolensky used probabilistic
polynomials over $\mathbb{F}_p$ to obtain exponential circuit size
lower bounds for $\mathsf{MAJORITY}$ and $\Mod q$, where $q\ne p$
is prime~\cite{razborov_1987,10.1145/28395.28404}.
Kopparty and Srinivasan refined the degree bound for
$p=2$~\cite{kopparty2018certifying}.
For a proof covering every prime $p$, see Oliveira and
Santhanam~\cite{oliveira_et_al:LIPIcs.CCC.2015.124}.

To study $\ACC$, one needs an approximation model that can handle
modular gates with different prime factors in the modulus.
Bhrushundi, Hosseini, Lovett, and Rao proposed torus polynomials for
this purpose and proved that $\ACC$ circuits have low-degree torus
approximations~\cite{bhrushundi_torus_2019}.
A torus polynomial $P$ is a real polynomial evaluated modulo one.
It approximates a Boolean function $f\colon\{0,1\}^n\to\{0,1\}$
by satisfying
 \begin{align*}
 \|P(x)-f(x)/2\|_{\mathbb{T}}&\le\epsilon
 \qquad\forall x\in\{0,1\}^n,
 \end{align*}
where $\mathbb{T}=\mathbb{R}/\mathbb{Z}$ and
$\|a\|_{\mathbb{T}}=\min_{z\in\mathbb{Z}}|a-z|$.
The factor $1/2$ keeps the two Boolean values distinct modulo one.
For a real-valued function $g$ on the Boolean cube, we write
$\tdeg(g)$ for the minimum degree of a polynomial $P$ satisfying
$\|P(x)-g(x)\|_{\mathbb{T}}\le\epsilon$ on every input.

We prove two torus approximation bounds. The first depends on the
Fourier $\ell_1$ norm and determines the degree of the AND function
up to constant factors at inverse-polynomial error. The second
depends on circuit size and depth, with explicit dependence on the
number of distinct prime factors of the modulus.

\paragraph{Fourier bounds and the AND function.}
We identify $\{0,1\}$ with $\mathbb{F}_2$. For
$f\colon\mathbb{F}_2^n\to\mathbb{R}$, let $\fnorm{f}$ be the sum of the
absolute values of its normalized Fourier coefficients.
\begin{theorem}
\label{thm:f1}
For every $f\colon\mathbb{F}_2^n\to\mathbb{R}$ and
$\epsilon\in(0,1/2)$,
\[
 \tdeg(f)=O\left(\log\frac{n(1+\fnorm{f})}{\epsilon}\right).
\]
\end{theorem}
Since $\fnorm{\AND_n}=1$, Corollary~\ref{cor:and} gives
\[
 \tdeg(\AND_n/2)=O(\log(n/\epsilon)).
\]
For $\epsilon=n^{-c}$ with fixed $c>0$, this matches the
$\Omega(\log(1/\epsilon))$ lower bound of Krishan and Vishwanathan
up to constant factors~\cite[Theorem~6]{krishan_et_al:LIPIcs.ITCS.2026.88}.
In particular, the degree at error $1/(20n)$ is $\Theta(\log n)$,
which refutes Conjecture~5 of Krishan and
Sarma~\cite{krishan_et_al:LIPIcs.MFCS.2026.62}.
Applying the same bound to $\alpha\AND_n$, for any fixed
$\alpha\in(0,1)$, also refutes their
Conjecture~24~\cite{krishan_et_al:LIPIcs.MFCS.2026.62}.

\paragraph{Modular circuits.}
The Fourier $\ell_1$ norm can be exponential in $n$ even for
functions computed by linear-size, depth-two $\AC$
circuits~\cite{bruck_smolensky_1992}.
Thus Theorem~\ref{thm:f1} alone does not give a polylogarithmic degree
bound for all small circuits.
Our second result uses the circuit structure to bound the torus
approximation degree in terms of circuit size and depth.
Here $\ACm$ circuits have unbounded-fan-in $\AND$ and $\Mod m$
gates, together with $\mathsf{NOT}$ gates. A $\Mod m$ gate outputs
one exactly when the sum of its inputs is divisible by $m$.
We count input vertices as part of the circuit size and do not count
$\mathsf{NOT}$ gates toward either size or depth.
\begin{theorem}\label{thm:intro}
For every $m\ge2$, there is a constant $C_m\ge1$ depending only on
$m$ such that the following holds. Let $\mu$ be the number of distinct
prime factors of $m$ and set $\nu=\max\{1,\mu-1\}$.
For every $n\ge1$, every $n$-input $\ACm$ circuit $C$ of depth at most $d$ and size
at most $s\ge2$, and every $\epsilon\in(0,1/4)$,
\[
 \tdeg(C/2)\le C_m^d(\log s)^{\nu d}\log(1/\epsilon).
\]
\end{theorem}
The exponent of $\log s$ is $(\mu-1)d$ when $\mu\ge2$, and $d$ when
$m$ is a prime power.
The proof of Corollary~20 in~\cite{bhrushundi_torus_2019} gives
a degree bound of $(\log s)^{O_m(d^2)}(\log(1/\epsilon))^{O_m(d)}$.
Our bound makes the exponent of the size logarithm linear in depth
and the dependence on $\log(1/\epsilon)$ linear.

Table~\ref{tab:polynomial-approximations} compares this result with
classical polynomial approximation bounds and their refinements.
Here $s\ge\max\{n,2\}$ includes the inputs, $d\ge2$ is fixed, and
$\epsilon\in(0,1/4)$. The primes $p\ne q$ and the modulus $m$ are fixed,
and $\mu$ counts the distinct prime factors of $m$.

\begin{table}[!t]
\centering
\small
\setlength{\tabcolsep}{3pt}
\renewcommand{\arraystretch}{1.25}
\begin{tabularx}{\textwidth}{@{}
  >{\raggedright\arraybackslash}p{0.11\textwidth}
  >{\raggedright\arraybackslash}p{0.24\textwidth}
  >{\raggedright\arraybackslash}p{0.335\textwidth}
  >{\raggedright\arraybackslash}X@{}}
\toprule
Circuit class & Approximation & Degree upper bound & Circuit size lower bound \\
\midrule
$\AC$
  & Real polynomials\newline
    $L_2(U_n)$ error
  & $O_d((L+E)^d)$\newline
    \cite{10.1145/174130.174138}\par\smallskip
    $O_d(LE(L+E)^{d-2})$\newline
    \cite{HASTAD2001498}\par\smallskip
    $O_d(L^{d-1}E)$\newline
    \cite{tal:LIPIcs.CCC.2017.15}
  & $\mathsf{PARITY}_n$:\newline
    $\exp(\Omega_d(n^{1/d}))$\newline
    \cite{10.1145/174130.174138}\par\smallskip
    $\exp(\Omega_d(n^{1/(d-1)}))$\newline
    \cite{hastad_1986,HASTAD2001498,tal:LIPIcs.CCC.2017.15} \\
\addlinespace[6pt]
$\mathsf{AC}^0[p]$
  & $\mathbb{F}_p$ polynomials\newline
    Pointwise failure probability
  & $O_{p,d}((L+E)^d)$\newline
    \cite{razborov_1987,10.1145/28395.28404}\par\smallskip
    $O_{p,d}(L^{d-1}E)$\newline
    \cite{kopparty2018certifying,oliveira_et_al:LIPIcs.CCC.2015.124}
  & $\Mod{q}$, $\mathsf{MAJORITY}_n$:\newline
    $\exp(\Omega_{p,q,d}(n^{1/(2d)}))$\newline
    \cite{razborov_1987,10.1145/28395.28404}\par\smallskip
    $\exp(\Omega_{p,q,d}(n^{1/(2(d-1))}))$\newline
    \cite{kopparty2018certifying,oliveira_et_al:LIPIcs.CCC.2015.124} \\
\addlinespace[6pt]
$\ACm$\newline($\mu\ge2$)
  & Torus polynomials\newline
    $L_\infty$ torus error
  & $L^{O_m(d^2)}E^{O_m(d)}$\newline
    \cite{bhrushundi_torus_2019}\par\smallskip
    $O_{m,d}(L^{(\mu-1)d}E)$\newline
    This work (Theorem~\ref{thm:intro})
  & $\mathsf{MAJORITY}_n$:\newline
    An unconditional superpolynomial bound is open~\cite{bhrushundi_torus_2019}. \\
\bottomrule
\end{tabularx}
\caption{Polynomial approximation bounds and circuit size lower bounds,
where $L=\log s$ and $E=\log(1/\epsilon)$.
The lower bounds in the first two rows concern exact computation and
follow by taking a sufficiently small constant approximation error.}
\label{tab:polynomial-approximations}
\end{table}

The three rows use different approximation guarantees.
For $\AC$, one real polynomial $P$ satisfies
$\|P-f\|_{L_2(U_n)}\le\epsilon$, where $U_n$ is the uniform input
distribution. For $\mathsf{AC}^0[p]$, a random polynomial $P$ satisfies
$\Pr_P[P(x)\ne f(x)]\le\epsilon$ for every fixed input $x$.
For torus polynomials, one polynomial $P$ satisfies
$\max_x\|P(x)-f(x)/2\|_{\mathbb{T}}\le\epsilon$.

\paragraph{Conditional lower bounds for the majority function.}
A sufficiently strong torus degree lower bound for
$\mathsf{MAJORITY}$ would imply a superpolynomial $\ACC$ circuit
size lower bound~\cite{bhrushundi_torus_2019}.
Conjecture~5 of~\cite{bhrushundi_torus_2019} states that every
torus polynomial approximating
$\mathsf{MAJORITY}_n/2$ with error $1/(20n)$ has degree
$\Omega(\sqrt{n/\log n})$. This bound is proved for symmetric
torus polynomials in~\cite[Corollary~23]{bhrushundi_torus_2019}.

Assume this conjecture and fix $m\ge2$ and $d\ge1$. If a depth-$d$
$\ACm$ circuit of size $s$ computes $\mathsf{MAJORITY}_n$, then the
earlier degree bound of~\cite{bhrushundi_torus_2019} at error $1/(20n)$ gives
\[
 \sqrt{\frac{n}{\log n}}
 \le O_{m,d}\!\left((\log s)^{O_m(d^2)}(\log n)^{O_m(d)}\right).
\]
Since $s\ge n$, we can absorb all powers of $\log n$ into the power
of $\log s$. Solving for $s$ gives $s\ge\exp(n^{c_m/d^2})$ for some
$c_m>0$ depending only on $m$ and all sufficiently large $n$.

At the same error, Theorem~\ref{thm:intro} gives degree
$O_{m,d}((\log s)^{\nu d}\log n)$. The conjecture then yields
\[
 s\ge\exp\!\left(\Omega_{m,d}\!\left(
       \left(\frac{n}{(\log n)^3}\right)^{1/(2\nu d)}
       \right)\right).
\]
Both bounds imply $\mathsf{MAJORITY}\notin\ACC$ under the conjecture.
For fixed $m$, the power of $n$ in the lower bound on $\log s$
improves from $c_m/d^2$ to $1/(2\nu d)$.

\paragraph{Proof ideas.}
For any modulus $q\ge2$, let $\delta_q(z)$ be one if
$z\equiv0\pmod q$ and zero otherwise.
Both proofs use approximations to weighted zero tests
$\alpha\delta_q(G(x))$, where $\alpha$ is real and $G$ is an integer
polynomial. When $q=p^e$ is a prime power, a finite-difference
argument represents $\delta_q/p^k$ modulo one by a rational
polynomial of degree at most $(q-1)k$.
For $0<\eta<1/2$, choose $k=\lceil\log_p(1/(2\eta))\rceil$ and round
$\alpha$ to the nearest integer multiple of $p^{-k}$.
Composing the resulting polynomial with $G$ gives error at most
$\eta$ and degree $O_q(t\log(1/\eta))$, where $t=\deg G$.
This bound is independent of $\alpha$ and of the coefficients of $G$.

For the Fourier bound, the identity
 \begin{align*}
 (-1)^{\sum_i\gamma_i x_i}
 &=2\delta_2\left(\sum_i\gamma_i x_i\right)-1
 \qquad \forall\gamma,x\in\{0,1\}^n
 \end{align*}
relates parity characters to zero tests.
If $f$ has $r\ge1$ nonzero Fourier coefficients, rounding each to
the nearest integer multiple of $2^{-k}$ gives real-valued error
at most $r2^{-k-1}$ on every input.
Taking $k=\lfloor\log(r/\epsilon)\rfloor$ bounds this error by
$\epsilon$, and the representation above then gives torus degree
$O(\log(r/\epsilon))$.
For a general $f$, Fourier sampling gives a weighted sum of
$O(1+n\fnorm{f}^2/\epsilon^2)$ characters that approximates $f$
with real-valued error $\epsilon/2$ on every
input~\cite{bruck_smolensky_1992,abf2014}.
Approximating this sum with torus error $\epsilon/2$ proves
Theorem~\ref{thm:f1}.

For circuits, write $m=\prod_{i=1}^{\mu}p_i^{e_i}$ and
$q_i=p_i^{e_i}$. The Chinese remainder theorem gives
$\delta_m(G(x))=\prod_{i=1}^{\mu}\delta_{q_i}(G(x))$.
We introduce one factor at a time. Multiplication by a Boolean
zero test preserves the previous torus error.
We then expand the current polynomial into monomials and apply
the prime-power bound to each coefficient times the new zero test.
Multiplying these approximations by the corresponding Boolean
monomials and summing them completes the step.
For integer affine $G$, the resulting approximation to
$\alpha\delta_m(G(x))$ has error $\eta$ and degree
\[
 O_m\bigl((\log(n+1))^{\mu-1}\log(1/\eta)\bigr).
\]

Section~\ref{sec:acc-torus} uses this bound in an induction on
circuit depth. Replacing gates whose inputs are original input
literals or constants by independent variables reduces the depth.
After approximating the smaller-depth circuit by a degree-$D$
torus polynomial, we substitute the gate functions back into its
monomials. Each monomial becomes a product of affine modular zero
tests. Character orthogonality expresses this product as
\[
 \prod_{j=1}^u\delta_m(G_j(x))
 =\E_{\lambda\in\{0,\ldots,m-1\}^u}
 \left[\cos\left(\frac{2\pi}{m}
                  \sum_{j=1}^u\lambda_jG_j(x)\right)\right].
\]
We sample monomials and characters jointly.
Reducing the polynomial coefficients modulo one bounds their
absolute sum by $(s+1)^D$. Concentration and a union bound give a
sample average that works on every input, with the logarithm of the
sample count bounded by $O((D+1)\log s+\log(1/\epsilon))$.
Each sampled cosine is a linear combination of $m$ affine zero
tests, one for each residue class. Applying the bound above to
each weighted test, with errors summing to at most $\epsilon/3$,
gives degree
\[
 O_m\bigl((\log s)^{\mu-1}
                ((D+1)\log s+\log(1/\epsilon))\bigr).
\]
Allocating the other two thirds of the error to the induction
hypothesis and sampling closes the induction with exponent $\mu d$.
The samples are fixed once for all inputs, producing a single
approximating polynomial.

For $\mu\ge2$, the improvement to $(\mu-1)d$ uses a circuit
decomposition described in Appendix~\ref{app:refined-circuit-bound}.
We process the factors $q_1,\ldots,q_\mu$ in order
at each circuit height. Reversing this order at alternate heights,
following Beigel and Tarui
\cite[Section~5.2, p.~789]{beigel_tarui_1991}, lets us merge
adjacent maps with the same modulus.
The composition has at most $(\mu-1)d+1$ Boolean maps,
each associated with a single prime-power modulus.
Its final map has a constant-degree polynomial representation
modulo that modulus and hence has torus approximation degree
$O_m(\log(1/\epsilon))$ after division by two.
Each of the remaining at most $(\mu-1)d$ substitutions contributes
one factor of $O_m(\log s)$, which gives Theorem~\ref{thm:intro}.

\paragraph{Application to quantum computation.}
Low torus degree gives upper bounds on the number of qubits needed
for $\NMQC$, non-adaptive measurement-based quantum computation with
linear side-processing. Through periodic Fourier representations,
our circuit bound gives quasipolynomial qubit upper bounds for
$\ACC$ at inverse-polynomial error. Fourier sampling also gives
$O(1+n\fnorm{f}^2/\epsilon)$ qubits for a Boolean function $f$, and
hence $O(n/\epsilon)$ qubits for the AND function, with error
probability at most $\epsilon$ on every input.
These constructions use fixed input parities and fixed local
measurements on a GHZ state.

The connection with periodic Fourier representations follows from
Werner and Wolf~\cite{werner_wolf_2001}, Hoban, Campbell, Loukopoulos,
and Browne~\cite{Hoban_2011}, and Mori~\cite{mori_periodic_2018}.
Their results characterize the optimum for each fixed input
distribution. Together with the minimax theorem, this gives
the worst-case error attainable by randomized periodic Fourier
representations, allowing both the parities and the coefficients
to depend on the random choice.
For $\ACC$, our torus approximations yield fixed periodic Fourier
representations that work on every input.
We also convert a periodic Fourier representation into a torus
approximation whose degree is logarithmic in its sparsity and in the
inverse additional error.

\paragraph{Organization.}
Section~\ref{sec:related-work} reviews related work.
Section~\ref{sec:preliminaries} defines the circuit
classes and torus approximation degree.
Section~\ref{sec:modular-approximation} proves the prime-power
representation and approximation lemmas and extends the approximation
bound to composite moduli.
Section~\ref{sec:fourier} proves the Fourier bounds and the corollary
for the AND function.
Section~\ref{sec:acc-torus} proves the circuit bound with exponent
$\mu d$. Section~\ref{sec:nmqc} relates torus degree to periodic
Fourier sparsity and gives the quantum constructions.
Section~\ref{sec:conclusion} summarizes the results and methods.
Appendix~\ref{app:refined-circuit-bound} proves
Theorem~\ref{thm:intro} for $\mu\ge2$.
Appendix~\ref{app:functional-degree} relates functional degree to
polynomial representation and records the exact degree
for prime-power zero tests.

\section{Related work}
\label{sec:related-work}
\paragraph{Polynomial representations of $\ACC$.}
Beigel and Tarui~\cite{beigel_tarui_1994} proved that every $\ACC$
function has a quasipolynomial-size $\mathsf{SYM}\circ\AND$
representation with polylogarithmic bottom fan-in.
Equivalently, $f(x)=h(P(x))$, where $P$ is an integer polynomial of
polylogarithmic degree with quasipolynomially bounded coefficients,
and $h\colon\mathbb{Z}\to\{0,1\}$ is arbitrary.
For fixed $m$, Chen and
Papakonstantinou~\cite{chen_papakonstantinou_2019} improved the size
bound for depth-$d$, size-$s$ $\ACm$ circuits from
$2^{(\log s)^{2^{O_m(d)}}}$ to $2^{(\log s)^{O_m(d)}}$.
Converting such a representation to a torus approximation also
requires control of the decoding function $h$.

Green, K{\"o}bler, Regan, Schwentick, and Tor{\'a}n~\cite{green_etal_1995}
gave a representation in which
a specified bit of an integer polynomial equals the circuit output.
The construction in~\cite[Theorem~19 and Corollary~20]{bhrushundi_torus_2019}
uses a strengthened form with zeros on both sides of that bit.
Dividing the polynomial by an appropriate power of two then gives
a torus approximation, with the degree bound discussed in the
introduction.

Our direct induction in Section~\ref{sec:acc-torus} uses the
character expansion from Chen and Papakonstantinou's linearization
of products of modular zero tests~\cite{chen_papakonstantinou_2019}.
They turn this expansion into an integer congruence modulo an
auxiliary coprime modulus.
We take its real part and sample monomials and characters jointly.
Approximating the resulting weighted zero tests gives the bound
with exponent $\mu d$ and linear dependence on $\log(1/\epsilon)$.
For $\mu\ge2$, the further improvement to $(\mu-1)d$ uses the
circuit decomposition in Appendix~\ref{app:refined-circuit-bound}.

\paragraph{Torus degree of the AND function.}
Krishan and Vishwanathan asked whether the known
$O(\log^2(n/\epsilon))$ upper bound for the AND function could be
improved to match their lower bound at inverse-polynomial
error~\cite[Lemma~17 and Open Problem~4]{krishan_et_al:LIPIcs.ITCS.2026.88}.
Krishan and Sarma formulated their degree conjectures as a route
to size lower bounds for certain depth-three modular
circuits~\cite{krishan_et_al:LIPIcs.MFCS.2026.62}.

Krishan and Vishwanathan also prove that every symmetric torus
polynomial approximating $\AND_n/2$ with error $1/(20n)$ has degree
$\Omega(\sqrt{n/\log n})$~\cite[Theorem~7]{krishan_et_al:LIPIcs.ITCS.2026.88}.
Our construction approximates $\AND_n/2$ with error $\epsilon$
using a torus polynomial of degree $O(\log(n/\epsilon))$ that need
not be symmetric.

\paragraph{Fourier sampling.}
Bruck and Smolensky~\cite{bruck_smolensky_1992} used Fourier sampling
to construct sparse polynomial threshold representations.
See also Siu and Bruck~\cite{siu_bruck_1991}.
The uniform real-valued approximation bound used in
Theorem~\ref{thm:f1} is stated in O'Donnell's
Theorem~5.12, with historical notes in Chapter~5~\cite{abf2014}.
We combine it with the conversion from Fourier sparsity to torus
degree in Theorem~\ref{thm:f0}.
Sampling is also used in~\cite[Lemma~14]{bhrushundi_torus_2019}
to convert probabilistic finite-field approximations into
one torus polynomial that approximates the target on every input.

\paragraph{Periodic Fourier representations.}
Periodic Fourier representations give GHZ-state implementations of
$\NMQC$~\cite{Hoban_2011,mori_periodic_2018}. In the error-free
setting, the minimum periodic Fourier sparsity of a nonconstant
Boolean function equals the minimum number of qubits.
For a nonconstant Boolean function $f$ on $n$ bits,
Mori~\cite[Construction~2 and Theorem~1]{mori_periodic_2018} gives
\[
 2^d-1\le\mathsf{pfs}_0(f)\le n^d,
 \qquad d=\deg_{\mathbb{F}_2}(f),
\]
where $\mathsf{pfs}_0(f)$ is the minimum number of nonconstant
parity characters in an exact periodic Fourier representation.
Consequently, a Boolean function family has polylogarithmic
$\mathbb{F}_2$-degree if and only if it has quasipolynomial periodic
Fourier sparsity in the error-free setting.
For bounded error, the same correspondence characterizes the optimum
average error under a fixed input distribution. Our pointwise
approximation bounds instead provide a single periodic Fourier
polynomial that works on every input.

\section{Preliminaries}
\label{sec:preliminaries}
\subsection{Notation}
Throughout the paper, $\log$ denotes the logarithm to base two, while
$\exp$ denotes the natural exponential function.
We write $[n]=\{1,\ldots,n\}$.
All polynomial degrees are ordinary total degrees unless explicitly
called functional degrees. On the Boolean cube, we multilinearize
polynomials without changing their values or increasing their degree.

\subsection{Circuit classes}
\label{sec:circuits}
An $\mathsf{AC}$ circuit consists of unbounded-fan-in $\mathsf{AND}$
gates and $\mathsf{NOT}$ gates. Constants are allowed as inputs to
gates. Its \term{depth} is the maximum number
of non-$\mathsf{NOT}$ gates on a directed path in the circuit.
Its \term{size} is the number of non-$\mathsf{NOT}$ gates plus the
number of input bits. Allowing $\mathsf{OR}$ gates gives the same model,
by De Morgan's laws.
We use a size upper bound $s\ge2$ in logarithmic expressions.
For a circuit of size one, we take $s=2$.

\begin{definition}[The class $\AC$]
A family of Boolean functions
$(f_n\colon\{0,1\}^n\to\{0,1\})_{n\ge1}$ belongs to $\AC$ if there
are nonnegative integers $c,d,k$ and $\mathsf{AC}$ circuits $(C_n)_{n\ge1}$
such that $C_n$ computes $f_n$, has depth at most $d$, and has size at
most $cn^k$, for every $n\ge1$.
\end{definition}

For $m\ge2$, a $\Mod{m}$ gate is defined by
\[
 \Mod{m}(x_1,\ldots,x_n)=
 \begin{cases}
  1,&x_1+\cdots+x_n\equiv0\pmod m,\\
  0,&\text{otherwise}.
 \end{cases}
\]
An $\mathsf{AC}[m]$ circuit may additionally use $\Mod{m}$ gates.
Its depth and size are defined in the same way as above, counting
$\Mod{m}$ gates as non-$\mathsf{NOT}$ gates.

\begin{definition}[The class $\ACm$]
A Boolean function family $(f_n)_{n\ge1}$ belongs to $\ACm$ if there
are nonnegative integers $c,d,k$ and $\mathsf{AC}[m]$ circuits $(C_n)_{n\ge1}$
such that $C_n$ computes $f_n$, has depth at most $d$, and has size at
most $cn^k$, for every $n\ge1$.
\end{definition}

\begin{definition}[The class $\ACC$]
The class $\ACC$ is the union
\[
 \ACC=\bigcup_{m\ge2}\ACm.
\]
In particular, the modulus is fixed throughout each circuit family.
\end{definition}

\subsection{Torus polynomials}
\label{sec:torus}
Let $\mathbb{T}=\mathbb{R}/\mathbb{Z}$.
\begin{definition}[Torus polynomial~\cite{bhrushundi_torus_2019}]
A function $P\colon\{0,1\}^n\to\mathbb{T}$ is a torus polynomial of
degree at most $d$ if there are real coefficients $\alpha_S$ such that
\[
 P(x)=\sum_{\substack{S\subseteq[n]\\|S|\le d}}
       \alpha_S\prod_{i\in S}x_i\pmod1.
\]
\end{definition}

\begin{definition}[Distance to zero on the torus]
For $\alpha\in\mathbb{R}$, let
\[
 \|\alpha\|_{\mathbb{T}}=\min_{z\in\mathbb{Z}}|\alpha-z|.
\]
This defines the distance to zero in $\mathbb{T}$.
\end{definition}

The distance between $\alpha,\beta\in\mathbb{T}$ is
$\|\alpha-\beta\|_{\mathbb{T}}$. It satisfies the triangle inequality
 \begin{align*}
 \|\alpha-\gamma\|_{\mathbb{T}}
 &\le \|\alpha-\beta\|_{\mathbb{T}}+\|\beta-\gamma\|_{\mathbb{T}}
 \qquad\forall \alpha,\beta,\gamma\in\mathbb{T},
 \end{align*}
which follows from the triangle inequality in $\mathbb{R}$ by
minimizing over integer shifts.

\begin{definition}[Torus approximation degree]
Let $f\colon\{0,1\}^n\to\mathbb{R}$ and $\epsilon>0$.
A torus polynomial $P$ \term{$\epsilon$-approximates} $f$ if
 \begin{align*}
 \|P(x)-f(x)\|_{\mathbb{T}}&\le\epsilon
 \qquad\forall x\in\{0,1\}^n.
 \end{align*}
The minimum degree of such a polynomial is denoted by $\tdeg(f)$.
We use degree zero for a constant approximating polynomial, including
the zero polynomial.
\end{definition}

This definition allows any real-valued target. For a Boolean function
$f$, the convention of~\cite{bhrushundi_torus_2019} corresponds to
$\tdeg(f/2)$. For this target, we restrict to $\epsilon\in(0,1/4)$:
the constant polynomial $P=1/4$ already gives error $1/4$ for every
Boolean function.

\section{Approximating modular zero tests}
\label{sec:modular-approximation}
For $m\ge2$ and $z\in\mathbb{Z}$, let $\delta_m(z)$ be one if
$z\equiv0\pmod m$ and zero otherwise.
For $q=p^e$, we first represent $\delta_q/p^k$ modulo one by a
polynomial of degree at most $(q-1)k$, then use coefficient rounding
to approximate $\alpha\delta_q(G(x))$.
We then extend the approximation bound to arbitrary moduli.

\begin{lemma}[Prime-power representation]\label{lem:prime-power-representation}
Let $q=p^e$, where $p$ is prime and $e\ge1$.
For every integer $k\ge1$, there is a polynomial
$Q_{q,k}\in\mathbb{Q}[z]$ of degree at most $(q-1)k$ such that
\[
 Q_{q,k}(z)=\frac{\delta_q(z)}{p^k}\pmod1
 \qquad\forall z\in\mathbb{Z}.
\]
\end{lemma}
\begin{proof}
On integer-valued $q$-periodic functions, let
$Tf(z)=f(z+1)$ and $\Delta=T-I$.
Since $q=p^e$, the identity $(1+X)^q=1+X^q$ over $\mathbb{F}_p$
shows that $p$ divides $\binom qj$ for $1\le j<q$.
As $T^q=I$, we obtain
\[
 \Delta^q=-\sum_{j=1}^{q-1}\binom qj\Delta^j
          =p\Delta A(\Delta)
\]
for some $A\in\mathbb{Z}[X]$.
Induction gives
\[
 \Delta^{k(q-1)+1}=p^k\Delta A(\Delta)^k\equiv0\pmod{p^k}.
\]
Put $d=(q-1)k$. For $0\le j\le d$, choose
$a_j\in\{0,\ldots,p^k-1\}$ such that
$a_j\equiv(\Delta^j\delta_q)(0)\pmod{p^k}$.
For $z\ge0$, expand $T^z=(I+\Delta)^z$ by the binomial theorem.
Since $\Delta^{d+1}$ annihilates every integer-valued $q$-periodic
function modulo $p^k$, all terms with powers of $\Delta$ greater
than $d$ vanish modulo $p^k$. Hence
\begin{align*}
 \delta_q(z)
 &= (T^z\delta_q)(0)\\
 &= \bigl((I+\Delta)^z\delta_q\bigr)(0)\\
 &\equiv \sum_{j=0}^d\binom zj(\Delta^j\delta_q)(0)\\
 &\equiv \sum_{j=0}^d a_j\binom zj\pmod{p^k}.
\end{align*}
The same expansion holds for negative $z$, with
$\binom zj=z(z-1)\cdots(z-j+1)/j!$ and $\binom z0=1$.
Indeed, nilpotence gives
\[
 T^{-1}=(I+\Delta)^{-1}
 \equiv\sum_{j=0}^d(-1)^j\Delta^j\pmod{p^k}.
\]
Raising this finite expansion to the power $-z$ gives the
generalized binomial expansion for $T^z$, again with all powers
above $d$ vanishing modulo $p^k$.

Define
\[
 Q_{q,k}(z)=\frac{1}{p^k}\sum_{j=0}^d a_j\binom zj.
\]
The congruence above implies
\[
 Q_{q,k}(z)-\frac{\delta_q(z)}{p^k}
 =\frac{\sum_{j=0}^d a_j\binom zj-\delta_q(z)}{p^k}
 \in\mathbb{Z}
 \qquad\forall z\in\mathbb{Z}.
\]
Thus $Q_{q,k}(z)=\delta_q(z)/p^k$ modulo one for every integer $z$.
Since $\binom zj$ is a rational polynomial of degree $j$,
we also have $\deg Q_{q,k}\le d=(q-1)k$.
\end{proof}

The exact degree of $Q_{q,k}$ is
\[
 (p-1)p^{e-1}k+p^{e-1}-1
 =(q-1)k-(p^{e-1}-1)(k-1).
\]
Appendix~\ref{app:functional-degree} proves this formula and its optimality.
The simpler bound $(q-1)k$ suffices for the approximation
arguments below.

\begin{remark}[Alternative representation via modulus amplification]
An alternative construction for the representation in
Lemma~\ref{lem:prime-power-representation} uses Beigel--Tarui's
modulus-amplifying polynomial~\cite{beigel_tarui_1994}.
Let $q=p^e$. For $z\ge0$, Lucas's theorem identifies
$\binom{z}{p^j}\bmod p$ with the $j$th base-$p$ digit of $z$
for $0\le j<e$.
Fermat's little theorem therefore gives
\[
 H_q(z)=\prod_{j=0}^{e-1}
       \left(1-\binom{z}{p^j}^{p-1}\right)
 \equiv\delta_q(z)\pmod p.
\]
Here $H_q$ is an integer-valued rational polynomial of degree $q-1$.
The congruence holds for every $z\in\mathbb{Z}$, since the binomial
coefficients in the product are $q$-periodic modulo $p$.
For $e=1$, this is the usual test $H_p(z)=1-z^{p-1}$.

Let $A_k\in\mathbb{Z}[X]$ be the modulus-amplifying polynomial
of degree $2k-1$, satisfying
\[
 A_k(u)\equiv b\pmod{p^k}
 \quad\text{whenever }u\equiv b\pmod p,\quad b\in\{0,1\}
\]
as stated in~\cite[Lemma~10]{bhrushundi_torus_2019}.
Then
\[
 \widetilde Q_{q,k}(z)=\frac{A_k(H_q(z))}{p^k}
 \equiv\frac{\delta_q(z)}{p^k}\pmod1,
 \qquad
 \deg\widetilde Q_{q,k}\le(q-1)(2k-1).
\]
Lemma~\ref{lem:prime-power-representation} gives the degree bound
$(q-1)k$. Since $(q-1)(2k-1)<2(q-1)k$, the construction above
also suffices for the asymptotic bounds in the subsequent sections.
\end{remark}

Rounding the weight to the nearest integer multiple of $p^{-k}$ gives the
approximation used below.

\begin{lemma}[Prime-power approximation]\label{lem:apxpe}
Let $q=p^e$, where $p$ is prime and $e\ge1$.
For every $\alpha\in\mathbb{R}$, $\epsilon\in(0,1/2)$, and integer
polynomial $G\in\mathbb{Z}[x_1,\ldots,x_n]$ of degree at most $t$,
\[
 \tdeg\bigl(\alpha\delta_q(G(x))\bigr)
 \le t(q-1)\log(1/\epsilon).
\]
\end{lemma}
\begin{proof}
Set $k=\lceil\log_p(1/(2\epsilon))\rceil\ge1$ and choose
$c\in\mathbb{Z}$ with
$|\alpha-c/p^k|\le1/(2p^k)\le\epsilon$.
By Lemma~\ref{lem:prime-power-representation},
$cQ_{q,k}(G(x))$ equals $(c/p^k)\delta_q(G(x))$ modulo one for
every Boolean $x$.
Since $\delta_q$ is Boolean, this approximates
$\alpha\delta_q(G(x))$ with torus error at most $\epsilon$.
Its degree is at most $t(q-1)k$, and
\[
 k\le1+\frac{\log(1/\epsilon)-1}{\log p}
 \le\log(1/\epsilon).
\]
Multilinearization does not increase the degree.
\end{proof}

The next lemma extends Lemma~\ref{lem:apxpe} to an arbitrary modulus.
We prove it by induction on the number of distinct prime factors.

\begin{lemma}[Composite-modulus approximation]\label{lem:apxm}
Let $m=\prod_{i=1}^{\mu}p_i^{e_i}$, where the $p_i$ are distinct primes,
and let $n,t\ge1$.
For every $\alpha\in\mathbb{R}$, $\epsilon\in(0,1/2)$, and integer
polynomial $G\in\mathbb{Z}[x_1,\ldots,x_n]$ of degree at most $t$,
\[
 \tdeg\bigl(\alpha\delta_m(G(x))\bigr)
 \le (m-1)t^\mu\bigl(\log(n+1)\bigr)^{\mu-1}\log(\mu/\epsilon).
\]
\end{lemma}
\begin{proof}
We induct on $\mu$. The case $\mu=1$ is Lemma~\ref{lem:apxpe}.
For $\mu\ge2$, put
\[
 q_i=p_i^{e_i},\qquad m'=\prod_{i=1}^{\mu-1}q_i,\qquad
 L=\log(n+1),\qquad \lambda=\log(\mu/\epsilon).
\]
The Chinese remainder theorem gives
\[
 \delta_m(G(x))=\delta_{q_\mu}(G(x))\delta_{m'}(G(x)).
\]
Apply the induction hypothesis with error $(\mu-1)\epsilon/\mu$.
We obtain a torus polynomial $P$ of degree at most
\[
 d=(m'-1)t^{\mu-1}L^{\mu-2}\lambda
\]
such that
\[
 \|P(x)-\alpha\delta_{m'}(G(x))\|_{\mathbb{T}}
 \le\frac{(\mu-1)\epsilon}{\mu}
 \qquad\forall x\in\{0,1\}^n.
\]
Multiplication by the Boolean value $\delta_{q_\mu}(G(x))$ preserves
this error bound. Thus
\[
 \|\delta_{q_\mu}(G(x))P(x)-\alpha\delta_m(G(x))\|_{\mathbb{T}}
 \le\frac{(\mu-1)\epsilon}{\mu}.
\]
Write $x_S=\prod_{i\in S}x_i$ and choose a real representative
$P(x)=\sum_{|S|\le d}\alpha_S x_S$.
There are at most $(n+1)^d$ terms in this sum.
By Lemma~\ref{lem:apxpe}, for each $S$ there is a torus polynomial
$P'_S$ of degree at most
\[
 t(q_\mu-1)\log\frac{\mu(n+1)^d}{\epsilon}
 =t(q_\mu-1)(\lambda+dL)
\]
that approximates $\alpha_S\delta_{q_\mu}(G(x))$ with error at most
$\epsilon/(\mu(n+1)^d)$.
Since $x_S\in\{0,1\}$, the polynomial
\[
 R(x)=\sum_{|S|\le d}P'_S(x)x_S
\]
approximates $\delta_{q_\mu}(G(x))P(x)$ with error at most
$\epsilon/\mu$. Consequently, $R$ approximates
$\alpha\delta_m(G(x))$ with error at most $\epsilon$.
Its degree is at most
\begin{align*}
 d+t(q_\mu-1)(\lambda+dL)
 &=d\bigl(1+t(q_\mu-1)L\bigr)+t(q_\mu-1)\lambda\\
 &\le dtq_\mu L+t(q_\mu-1)\lambda\\
 &\le\bigl((m'-1)q_\mu+(q_\mu-1)\bigr)
       t^\mu L^{\mu-1}\lambda\\
 &=(m-1)t^\mu L^{\mu-1}\lambda,
\end{align*}
where we used $t,L\ge1$.
\end{proof}

\section{Torus approximation from Fourier bounds}
\label{sec:fourier}
We identify $\{0,1\}$ with $\mathbb{F}_2$ and use the normalized
Fourier transform. For background, see~\cite{abf2014}.
For $\gamma,x\in\mathbb{F}_2^n$, let
$\ip{\gamma}{x}=\sum_{i=1}^n\gamma_i x_i$ in $\mathbb{F}_2$.
We also index characters by subsets $S\subseteq[n]$, identifying
$S$ with its indicator vector.

\begin{definition}[Fourier coefficients]
For $f\colon\mathbb{F}_2^n\to\mathbb{R}$, define
\[
 \widehat f(\gamma)=\E_{x\in\mathbb{F}_2^n}
       \bigl[f(x)(-1)^{\ip{\gamma}{x}}\bigr],
 \qquad \forall\gamma\in\mathbb{F}_2^n,
\]
where $x$ is uniform.
\end{definition}
Fourier inversion gives
\[
 f(x)=\sum_{\gamma\in\mathbb{F}_2^n}
        \widehat f(\gamma)(-1)^{\ip{\gamma}{x}}.
\]

\begin{definition}[Fourier sparsity]
The Fourier sparsity of $f\colon\mathbb{F}_2^n\to\mathbb{R}$ is
\[
 \suppsize{f}
 =\bigl|\{\gamma\in\mathbb{F}_2^n:\widehat f(\gamma)\ne0\}\bigr|.
\]
\end{definition}

We first bound torus degree by Fourier sparsity. Round the nonzero
Fourier coefficients to rational numbers with a common denominator
$2^k$. Since the characters take values in $\{\pm1\}$, the sum of
the coefficient errors bounds the error on every input.
Lemma~\ref{lem:prime-power-representation} then represents each
weighted character modulo one with degree at most $k-1$.

\begin{theorem}[Approximation from Fourier sparsity]\label{thm:f0}
For every $f\colon\mathbb{F}_2^n\to\mathbb{R}$ and
$\epsilon\in(0,1/2)$,
\[
 \tdeg(f)\le
 \log\frac{\max\{1,\suppsize{f}\}}{2\epsilon}.
\]
\end{theorem}
\begin{proof}
Let $r=\suppsize{f}$. If $r=0$, then $f=0$ and the claim is immediate.
Otherwise, set $k=\lfloor\log(r/\epsilon)\rfloor\ge1$.
For each $\gamma$, choose $c_\gamma\in\mathbb{Z}$ such that
\[
 \left|\widehat f(\gamma)-\frac{c_\gamma}{2^k}\right|
 \le2^{-k-1},
\]
taking $c_\gamma=0$ whenever $\widehat f(\gamma)=0$. Define
\[
 \widetilde f(x)=\sum_\gamma\frac{c_\gamma}{2^k}
                    (-1)^{\ip{\gamma}{x}}.
\]
Fourier inversion gives
\[
 |f(x)-\widetilde f(x)|\le r2^{-k-1}\le\epsilon
 \qquad\forall x\in\mathbb{F}_2^n.
\]
Moreover,
\[
 \widetilde f(x)
 =\sum_\gamma\frac{c_\gamma}{2^{k-1}}
       \delta_2\left(\sum_{i=1}^n\gamma_i x_i\right)
       -\sum_\gamma\frac{c_\gamma}{2^k}.
\]
If $k\ge2$, Lemma~\ref{lem:prime-power-representation} gives a torus polynomial
representation of $\widetilde f$ of degree at most $k-1$.
If $k=1$, the delta-function terms have integer coefficients, so
$\widetilde f$ is constant modulo one and the same bound holds.
In either case,
\[
 \tdeg(f)\le k-1\le\log\frac{r}{2\epsilon}.
\]
\end{proof}

\begin{definition}[Fourier $\ell_1$ norm]
The Fourier $\ell_1$ norm of $f\colon\mathbb{F}_2^n\to\mathbb{R}$ is
\[
 \fnorm f=\sum_{\gamma\in\mathbb{F}_2^n}|\widehat f(\gamma)|.
\]
\end{definition}

A function may have many nonzero Fourier coefficients but a small
Fourier $\ell_1$ norm. Sampling reduces the support before we apply
Theorem~\ref{thm:f0}. The argument of Bruck and
Smolensky~\cite{bruck_smolensky_1992}, as presented in
O'Donnell~\cite[Theorem~5.12]{abf2014}, approximates such a function
uniformly in the ordinary real-valued distance. We include the
argument to keep track of the error and the number of samples.

\begin{fourierrestatement}[Restated]
For every $f\colon\mathbb{F}_2^n\to\mathbb{R}$ and
$\epsilon\in(0,1/2)$,
\[
 \tdeg(f)=O\left(\log\frac{n(1+\fnorm{f})}{\epsilon}\right).
\]
\end{fourierrestatement}
\begin{proof}
The claim is immediate when $f=0$. Otherwise, choose $\gamma$ with
probability $|\widehat f(\gamma)|/\fnorm{f}$ and define
\[
 X_\gamma(x)=\fnorm{f}\,
 \sgn(\widehat f(\gamma))(-1)^{\ip{\gamma}{x}}.
\]
Then $|X_\gamma(x)|=\fnorm{f}$ and
$\E_\gamma[X_\gamma(x)]=f(x)$.
Set
\[
 M=\left\lceil\frac{8(n+2)\fnorm{f}^2}{\epsilon^2}\right\rceil
\]
and choose $\gamma^{(1)},\ldots,\gamma^{(M)}$ independently from this
distribution. For each fixed $x$, Hoeffding's
inequality~\cite{hoeffding_1963} gives
\begin{align*}
 \Pr\left[
 \left|\frac1M\sum_{j=1}^M X_{\gamma^{(j)}}(x)-f(x)\right|
 >\frac{\epsilon}{2}\right]
 &\le2\exp\left(-\frac{\epsilon^2 M}{8\fnorm{f}^2}\right)\\
 &\le 2^{-(n+1)}.
\end{align*}
A union bound over all $2^n$ inputs therefore shows that some choice
of samples satisfies
\[
 \left|\frac1M\sum_{j=1}^M X_{\gamma^{(j)}}(x)-f(x)\right|
 \le\frac{\epsilon}{2}
 \qquad\forall x\in\{0,1\}^n.
\]
Fix such a choice and let $g(x)=M^{-1}\sum_{j=1}^M X_{\gamma^{(j)}}(x)$.
Its Fourier sparsity is at most $M$.
Theorem~\ref{thm:f0} gives an $\epsilon/2$-approximating torus polynomial
for $g$ of degree at most $\log(M/\epsilon)$.
This polynomial approximates $f$ with error at most $\epsilon$.
Finally,
\[
 M\le1+\frac{8(n+2)\fnorm{f}^2}{\epsilon^2}
\]
implies
$\log(M/\epsilon)=O(\log(n(1+\fnorm{f})/\epsilon))$.
\end{proof}

Since $\fnorm{\AND_n}=1$, we immediately obtain the following bound.
\begin{corollary}\label{cor:and}
For every $n\ge1$ and $\epsilon\in(0,1/4)$,
\[
 \tdeg(\AND_n/2)=O(\log(n/\epsilon)).
\]
\end{corollary}

\begin{remark}
More generally, for every fixed $\alpha\in(0,1)$,
$\fnorm{\alpha\AND_n}=\alpha$, so Theorem~\ref{thm:f1} gives
\[
 \tdeg(\alpha\AND_n)=O(\log(n/\epsilon)).
\]
For the target $\AND_n/2$ and $\epsilon=n^{-c}$ with fixed $c>0$,
Corollary~\ref{cor:and} matches the known lower bound up to constant
factors~\cite{krishan_et_al:LIPIcs.ITCS.2026.88}.
This bound does not determine the optimal degree at constant error.
\end{remark}

\begin{remark}[Explicit construction for the AND function]
For the AND function, small-bias spaces give a deterministic alternative to
Fourier sampling. Let $S$ be an $\eta$-biased multiset in
$\mathbb{F}_2^n$, meaning that
$|\E_{a\in S}[(-1)^{\ip{a}{y}}]|\le\eta$ for every nonzero
$y\in\mathbb{F}_2^n$.
Explicit constructions give $|S|=\operatorname{poly}(n/\eta)$
\cite{naor_naor_1993,alon_etal_1992}.
Then
\[
 P_S(x)=\frac{1}{2|S|}\sum_{a\in S}(-1)^{\ip{a}{\mathbf{1}-x}}
 \qquad\text{satisfies}\qquad
 \|P_S-\AND_n/2\|_\infty\le\eta/2.
\]
Indeed, $P_S(\mathbf{1})=1/2$, and the small-bias condition applies
to every other input.
For $0<\epsilon<1/4$, take $\eta=\epsilon$ and apply the
coefficient-rounding construction in Theorem~\ref{thm:f0} with
error $\epsilon/2$.
Since $P_S$ has Fourier sparsity at most $|S|$, this gives an
explicit $\epsilon$-approximating torus polynomial for $\AND_n/2$
of degree $O(\log(n/\epsilon))$.
\end{remark}

The Fourier $\ell_1$ norm can be exponential in the number of inputs
even for small depth-two circuits~\cite{bruck_smolensky_1992}.
For example,
\[
 f(x,y)=\prod_{j=1}^r(1-x_jy_j)
 \qquad\text{has}\qquad \fnorm{f}=(3/2)^r.
\]
This function has an $\AC$ circuit of depth two and size $O(r)$.
Each factor has Fourier $\ell_1$ norm $3/2$, and the norms multiply
because the factors depend on disjoint pairs of variables.
Thus Theorem~\ref{thm:f1} alone does not give a polylogarithmic bound
for all small circuits. The next section uses the gate structure to
obtain such a bound.

\section{From modular circuits to torus polynomials}
\label{sec:acc-torus}
We prove a circuit approximation bound with exponent $\mu d$
directly by induction on depth, using Lemma~\ref{lem:apxm}.
At each step, we substitute gates whose inputs are original input
literals or constants.
For $\mu=1$, this proves Theorem~\ref{thm:intro}.
For $\mu\ge2$, Appendix~\ref{app:refined-circuit-bound} gives the
decomposition argument that improves the exponent to $(\mu-1)d$.

\begin{proposition}\label{prop:basic-circuit-bound}
For every $m\ge2$, there is a constant $A_m\ge1$ depending only on $m$
such that the following holds.
Let $\mu$ be the number of distinct prime factors of $m$.
For every $n$-input $\ACm$ circuit $C$ of depth at most $d$ and size
at most $s\ge2$, and every $\epsilon\in(0,1/4)$,
\[
 \tdeg(C/2)\le A_m^d(\log s)^{\mu d}\log(1/\epsilon).
\]
\end{proposition}
\begin{proof}
We will prove the bound
\[
 \tdeg(C/2)\le B_m^d(\log(s+1))^{\mu d}\log(1/\epsilon),
\]
where $B_m\ge1$ is a constant depending only on $m$, to be chosen below.
Since $\log(s+1)\le2\log s$ for $s\ge2$, this implies the stated
bound with $A_m=2^\mu B_m$.
Let $L=\log(s+1)$ and $E=\log(1/\epsilon)$, so that $L,E\ge1$.
Replace each $\AND$ gate by a $\NOR$ gate with all its inputs
negated. This does not change the circuit size or depth.
We induct on $d$.
If $d=0$, the output is a constant or an input literal, so $C/2$ is
represented exactly by a torus polynomial of degree at most one.

Assume $d\ge1$ and that this bound holds for depth at most $d-1$.
We divide the induction step into three parts.

\paragraph{Applying the induction hypothesis.}
Let $k$ be the number of gates all of whose inputs are original input
literals or constants, and let their outputs be
$h_{n+1}(x),\ldots,h_{n+k}(x)$.
Set $r=n+k$ and $h_a(x)=x_a$ for $1\le a\le n$.
Replace these $k$ gates by new independent input variables, keeping
the original inputs, to obtain a circuit $C'$ satisfying
\[
 C(x)=C'(h_1(x),\ldots,h_r(x)).
\]
Consider any path through the remaining gates. Its first gate was not
removed, so in $C$ one of its inputs is, up to negation, the output of
another gate; following such inputs downward ends at a removed gate.
Prepending this chain to the path adds at least one gate in the
original circuit. Thus $C'$ has depth at most $d-1$.
It has $r$ inputs and at most $s-n-k$ non-$\mathsf{NOT}$ gates.
Its size is therefore at most
\[
 r+(s-n-k)=s.
\]
Thus the same size bound $s$ can be used in the induction hypothesis.

By that hypothesis, there is a torus polynomial $P$ satisfying
\[
 \|C'(y)/2-P(y)\|_{\mathbb{T}}\le\epsilon/3
 \qquad\forall y\in\{0,1\}^r,
\]
whose degree $D$ satisfies
\begin{equation}\label{eq:basic-circuit-degree}
 D\le B_m^{d-1}L^{\mu(d-1)}\log(3/\epsilon).
\end{equation}
Each Boolean monomial takes values in $\{0,1\}$, so changing a
coefficient by an integer does not change the torus-valued function.
We may therefore write
\[
 P(y)=\sum_{S\in\mathcal U}\alpha_S\prod_{a\in S}y_a\pmod1,
 \qquad
 \mathcal U=\{S\subseteq[r]:|S|\le D\},
\]
with $\alpha_S\in[-1/2,1/2)$.
Let
\[
 K=|\mathcal U|,\qquad A=\sum_{S\in\mathcal U}|\alpha_S|.
\]
Then $K\le(r+1)^D\le2^{DL}$ and $A\le K/2$.
If $A=0$, the zero polynomial already gives the desired approximation,
so assume $A>0$.
Define real-valued functions
\[
 b_S(x)=\prod_{a\in S}h_a(x),\qquad
 Q(x)=\sum_{S\in\mathcal U}\alpha_S b_S(x).
\]
Substitution into $P$ gives
\begin{equation}\label{eq:basic-circuit-error-induction}
 \|C(x)/2-Q(x)\|_{\mathbb{T}}\le\epsilon/3
 \qquad\forall x\in\{0,1\}^n.
\end{equation}

\paragraph{Replacing products of zero tests by a short sum.}
Every $b_S$ is a product of modular zero tests of integer affine
polynomials. To see this, write the input literals of a removed gate
as $\ell_j(x)\in\{x_i,1-x_i,0,1\}$ and use
\begin{align*}
 \NOR(\ell_1(x),\ldots,\ell_u(x))
 &=\prod_{j=1}^u\delta_m(\ell_j(x)),\\
 \Mod m(\ell_1(x),\ldots,\ell_u(x))
 &=\delta_m\left(\sum_{j=1}^u\ell_j(x)\right),\\
 x_i&=\delta_m(x_i-1).
\end{align*}
Thus, for some integer affine polynomials
$G_{S,1},\ldots,G_{S,u_S}$,
\[
 b_S(x)=\prod_{j=1}^{u_S}\delta_m(G_{S,j}(x)).
\]
By character orthogonality, as in the expansion
of~\cite{chen_papakonstantinou_2019},
\begin{align*}
 b_S(x)
 &=\E_{\lambda\in\{0,\ldots,m-1\}^{u_S}}
       \left[\exp\left(\frac{2\pi i}{m}
                 \sum_{j=1}^{u_S}\lambda_j G_{S,j}(x)\right)\right]\\
 &=\E_\lambda\left[\cos\left(\frac{2\pi}{m}G_{S,\lambda}(x)\right)\right],
\end{align*}
where $G_{S,\lambda}(x)=\sum_{j=1}^{u_S}\lambda_j G_{S,j}(x)$.

To approximate $Q$ by a short sum, we combine these expectations into
one and then approximate it by a sample average.
Sample an index $S\in\mathcal U$ with probability $|\alpha_S|/A$,
and conditional on $S$, sample $\lambda$ uniformly from
$\{0,\ldots,m-1\}^{u_S}$.
Define the random function
\[
 Z(x)=\sgn(\alpha_S)
       \cos\left(\frac{2\pi}{m}G_{S,\lambda}(x)\right).
\]
Then $|Z(x)|\le1$ and
\[
 A\E[Z(x)]
 =A\sum_{S\in\mathcal U}\frac{|\alpha_S|}{A}
       \sgn(\alpha_S)b_S(x)
 =Q(x).
\]
Take
\[
 M=\left\lceil\frac{18(s+1)A^2}{\epsilon^2}\right\rceil
\]
independent samples $Z_1,\ldots,Z_M$.
For each fixed input $x$, Hoeffding's inequality~\cite{hoeffding_1963} yields
\[
 \Pr\left[
 \left|\frac A M\sum_{j=1}^M Z_j(x)-Q(x)\right|>\epsilon/3
 \right]
 \le2\exp\left(-\frac{\epsilon^2 M}{18A^2}\right)<2^{-s}.
\]
Since $n\le s$, a union bound over all $2^n$ inputs has failure
probability strictly less than one.
We may therefore fix samples and define
\[
 \widetilde Q(x)=\frac A M\sum_{j=1}^M Z_j(x)
\]
so that
\begin{equation}\label{eq:basic-circuit-error-sampling}
 |\widetilde Q(x)-Q(x)|\le\epsilon/3
 \qquad\forall x\in\{0,1\}^n.
\end{equation}

\paragraph{Approximating the sampled terms.}
For every integer affine polynomial $G_{S,\lambda}$,
\[
 \cos\left(\frac{2\pi}{m}G_{S,\lambda}(x)\right)
 =\sum_{a=0}^{m-1}\cos\left(\frac{2\pi a}{m}\right)
       \delta_m(G_{S,\lambda}(x)-a).
\]
Consequently, $\widetilde Q$ is a linear combination of at most $Mm$
modular zero tests.
Apply Lemma~\ref{lem:apxm} to each weighted test with $t=1$ and
error $\epsilon/(3Mm)$, and sum the resulting polynomials.
This gives a torus polynomial $P_{\mathrm{new}}$ satisfying
\begin{equation}\label{eq:basic-circuit-error-approximation}
 \|\widetilde Q(x)-P_{\mathrm{new}}(x)\|_{\mathbb{T}}
 \le\epsilon/3
 \qquad\forall x\in\{0,1\}^n
\end{equation}
and
\[
 \deg P_{\mathrm{new}}
 \le(m-1)(\log(n+1))^{\mu-1}\log\frac{3\mu Mm}{\epsilon}.
\]
Combining \eqref{eq:basic-circuit-error-induction},
\eqref{eq:basic-circuit-error-sampling}, and
\eqref{eq:basic-circuit-error-approximation} by the triangle inequality proves that $P_{\mathrm{new}}$
$\epsilon$-approximates $C/2$.

It remains to bound its degree.
Since $A\le K/2$, we have $M\le5(s+1)K^2/\epsilon^2$.
Together with $K\le2^{DL}$, this gives
\[
 \deg P_{\mathrm{new}}
 =O_m\bigl(L^{\mu-1}((D+1)L+E)\bigr).
\]
Using \eqref{eq:basic-circuit-degree}, $\log(3/\epsilon)\le3E$,
and $B_m,L,E\ge1$, we obtain
\[
 \deg P_{\mathrm{new}}
 \le c_m B_m^{d-1}L^{\mu d}E
\]
for a constant $c_m$ depending only on $m$.
Choosing $B_m\ge\max\{1,c_m\}$ closes the induction.
\end{proof}

\section{Application to quantum computation}
\label{sec:nmqc}
\subsection{Periodic Fourier polynomials and average error}
We now use torus approximation degree and Fourier sampling to bound
the number of qubits needed for non-adaptive measurement-based
quantum computation with linear side-processing, denoted by $\NMQC$.
In $\NMQC$, the resource state, the two binary measurements available
at each qubit, and the input parities are fixed before the input is
given. Each input parity selects one of the two local measurements.
The measurement choices do not depend on measurement outcomes, and
the output is the parity of those outcomes.
We first state the correspondence with periodic Fourier polynomials
for average error. Applying the minimax theorem gives the optimal
worst-case error when shared classical randomness may choose the
protocol. Our upper bounds will use fixed representations.

Write $\ee(a)=\exp(2\pi i a)$ for the map from $\mathbb{T}$ to
the unit circle $S^1$.

\begin{definition}[Error and distance on the torus]
\label{def:torus-error-distance}
For $a,b\in\mathbb{T}$, define
\begin{align*}
 \operatorname{Err}(a,b)
 &=\frac{1-\cos(2\pi(a-b))}{2}
   =\sin^2(\pi(a-b)),\\
 \mathsf{D}(a,b)
 &=\frac{|\ee(a)-\ee(b)|}{2}=|\sin(\pi(a-b))|.
\end{align*}
\end{definition}
Thus $\operatorname{Err}=\mathsf{D}^2$, and $\mathsf{D}$ is a
translation-invariant metric on $\mathbb T$.

\begin{fact}\label{fact:torus-circle}
For all $a,b\in\mathbb{R}$,
\[
 2\|a-b\|_{\mathbb{T}}
 \le\mathsf{D}(a,b)
 \le\pi\|a-b\|_{\mathbb{T}}.
\]
\end{fact}
This follows from $2t\le\sin(\pi t)\le\pi t$ for $0\le t\le1/2$.

Let $\phi\colon\{0,1\}\to\{\pm1\}$ be given by $\phi(x)=1-2x$,
and let $\phi^{\otimes n}$ denote its coordinatewise application.

\begin{definition}[Periodic Fourier polynomial and sparsity]
\label{def:periodic-fourier}
We call a real multilinear polynomial
\[
 P(u)=\sum_{S\subseteq[n]}\alpha_S\prod_{i\in S}u_i,
 \qquad \alpha_S\in\mathbb R,
\]
a \term{periodic Fourier polynomial} when we interpret its values on
$\{\pm1\}^n$ modulo one.
Following~\cite{mori_periodic_2018}, its \term{periodic Fourier sparsity}
is the number of nonzero nonconstant monomials,
\[
 \mathsf{pfs}(P)
 =\bigl|\{S\subseteq[n]:S\ne\varnothing,\ \alpha_S\ne0\}\bigr|.
\]
\end{definition}

Werner and Wolf's optimization of full-correlation Bell expressions,
the correspondence in~\cite[Theorem~2]{Hoban_2011}, and its periodic
Fourier formulation in~\cite[Eq.~(4) and Section~1.2]{mori_periodic_2018}
give the following characterization for each fixed input distribution.

\begin{theorem}[{Average error and periodic Fourier sparsity~\cite{werner_wolf_2001,Hoban_2011,mori_periodic_2018}}]
\label{thm:average-error-nmqc}
Let $f\colon\{0,1\}^n\to\{0,1\}$ be a Boolean function, let
$\mathcal D$ be a probability distribution on $\{0,1\}^n$, and let
$r\ge1$ be an integer and $\epsilon\in[0,1]$.
There exists an $\NMQC$ computation using at most $r$ qubits with
average error at most $\epsilon$ under $\mathcal D$ if and only if
there exists a periodic Fourier polynomial $P$ with
$\mathsf{pfs}(P)\le r$ satisfying
\[
 \E_{x\sim\mathcal D}\left[
 \operatorname{Err}\bigl(f(x)/2,P(\phi^{\otimes n}(x))\bigr)
 \right]\le\epsilon.
\]
The equivalence also holds for fixed input parities, with $P$
restricted to the corresponding nonconstant monomials.
\end{theorem}

\paragraph{Worst-case error.}
Fix $r\ge1$, and let $\mathcal P_r$ be the set of periodic Fourier
polynomials with sparsity at most $r$.
For each input distribution $\mathcal D$,
Theorem~\ref{thm:average-error-nmqc} identifies the minimum average
error of an $\NMQC$ computation using at most $r$ qubits with the
minimum over $P\in\mathcal P_r$. The choice of input parities may
depend on $\mathcal D$.
The minimax theorem gives
\begin{align*}
 &\max_{\mathcal D}\min_{P\in\mathcal P_r}
   \E_{x\sim\mathcal D}\left[
    \operatorname{Err}\bigl(f(x)/2,P(\phi^{\otimes n}(x))\bigr)
   \right]\\
 &\qquad=\min_{\Lambda}\max_{x\in\{0,1\}^n}
   \E_{P\sim\Lambda}\left[
    \operatorname{Err}\bigl(f(x)/2,P(\phi^{\otimes n}(x))\bigr)
   \right],
\end{align*}
where $\Lambda$ ranges over probability distributions on
$\mathcal P_r$.
Thus an average-error guarantee for every input distribution is
equivalent to a randomized periodic Fourier representation with the
same error bound on every input. Each sampled polynomial has at most
$r$ nonconstant monomials, but their supports need not coincide.
Both the parities and the coefficients may be random.
Operationally, this permits shared classical randomness to choose
an $\NMQC$ protocol before the input is given.

For $\ACC$, the construction below gives a single fixed polynomial
with the required error bound on every input.

\begin{definition}[Periodic Fourier representation with error]
\label{def:periodic-fourier-error}
Let $f\colon\{0,1\}^n\to\{0,1\}$ be a Boolean function and let
$0\le\epsilon<1/2$. A periodic Fourier polynomial $P$ is a
\term{periodic Fourier representation} of $f$ with error at most
$\epsilon$ if
\[
 \operatorname{Err}\bigl(f(x)/2,P(\phi^{\otimes n}(x))\bigr)
 \le\epsilon
 \qquad\forall x\in\{0,1\}^n.
\]
We write $\pfs(f)$ for the minimum of $\mathsf{pfs}(P)$ over all
such polynomials.
\end{definition}

The upper bounds below construct fixed periodic Fourier
representations and hence give $\NMQC$ computations with fixed
input parities and fixed local measurement pairs.

\subsection{Qubit upper bounds and torus degree}
The GHZ construction gives a pointwise upper bound in terms of the
minimum sparsity of a fixed periodic Fourier representation.

\begin{proposition}[{\cite{Hoban_2011,mori_periodic_2018}}]
\label{prop:pfs-nmqc}
For every Boolean function $f$ and $0<\epsilon<1/2$, there is an
$\NMQC$ computation of $f$ that uses at most $\max\{1,\pfs(f)\}$ qubits
and has error probability at most $\epsilon$ on every input.
\end{proposition}

Fact~\ref{fact:torus-circle} bounds the error of a periodic Fourier
representation in terms of torus distance:
\[
 \operatorname{Err}(a,b)\le\pi^2\|a-b\|_{\mathbb{T}}^2.
\]
For $0<\epsilon<1/2$, the torus error $\sqrt{\epsilon}/\pi$ is less
than $1/4$.
Let $D=\mathsf{deg}_{\mathbb{T},\sqrt{\epsilon}/\pi}(f/2)$.
The affine change of variables $\phi^{-1}(u)=(1-u)/2$ does not
increase degree, so a degree-$D$ torus approximation yields a periodic
Fourier representation with at most $(n+1)^D$ nonconstant monomials.
Hence
\[
 \pfs(f)\le(n+1)^D.
\]
Together with Theorem~\ref{thm:intro}, this proves the following
corollary.

\begin{corollary}\label{cor:acc-nmqc}
Fix $m\ge2$.
If $f\colon\{0,1\}^n\to\{0,1\}$ is computed by an $\ACm$ circuit
of depth at most $d$ and size at most $s\ge2$, then, for every $0<\epsilon<1/2$,
it can be computed by $\NMQC$ with error probability at most
$\epsilon$ on every input using at most
\[
 2^{(\log s)^{O_m(d)}\log(1/\epsilon)}
\]
qubits.
\end{corollary}

The Fourier sampling argument used in the proof of
Theorem~\ref{thm:f1} also bounds periodic Fourier sparsity
directly~\cite{bruck_smolensky_1992,abf2014}.

\begin{proposition}\label{prop:fourier-nmqc}
For every $n\ge1$, every Boolean function
$f\colon\{0,1\}^n\to\{0,1\}$, and every $0<\epsilon<1/2$,
\[
 \pfs(f)=O\left(\frac{n\fnorm{f}^2}{\epsilon}\right).
\]
Consequently, $f$ can be computed by $\NMQC$ using
$O(1+n\fnorm{f}^2/\epsilon)$ qubits with error probability at most
$\epsilon$ on every input.
\end{proposition}
\begin{proof}
For $f=0$, we have $\pfs(f)=0$ and one qubit suffices.
Assume $f\ne0$, so that $\fnorm{f}\ge\max_x|f(x)|=1$.
The sampling argument in the proof of Theorem~\ref{thm:f1}, applied
to $f/2$ with deviation $\sqrt{\epsilon}/\pi$ and
$M=\lceil\pi^2(n+2)\fnorm{f}^2/(2\epsilon)\rceil$ samples, gives a
function $g\colon\{0,1\}^n\to\mathbb{R}$ such that
\[
 \|g-f/2\|_\infty\le\frac{\sqrt{\epsilon}}{\pi},
 \qquad
 \suppsize{g}\le M=O\left(\frac{n\fnorm{f}^2}{\epsilon}\right).
\]
Set
\[
 P(u)=\sum_{S\subseteq[n]}\widehat g(S)\prod_{i\in S}u_i.
\]
Then $P(\phi^{\otimes n}(x))=g(x)$, and on every input,
\[
 \operatorname{Err}\bigl(f(x)/2,P(\phi^{\otimes n}(x))\bigr)
 =\sin^2\bigl(\pi(f(x)/2-g(x))\bigr)
 \le\pi^2|f(x)/2-g(x)|^2\le\epsilon.
\]
Thus $P$ is a periodic Fourier representation with the claimed
sparsity. The qubit bound follows from
Proposition~\ref{prop:pfs-nmqc}.
\end{proof}

Since $\fnorm{\AND_n}=1$, Proposition~\ref{prop:fourier-nmqc} gives
$\mathsf{pfs}_{\epsilon}(\AND_n)=O(n/\epsilon)$ and an
$O(n/\epsilon)$-qubit $\NMQC$ computation of $\AND_n$.
Both the parity family and the coefficients are fixed in this
construction.
If the input parities may be chosen at random, Smolensky's construction,
as discussed by Mori~\cite[Section~1.4]{mori_periodic_2018}, gives an
$O(1/\epsilon)$-qubit $\NMQC$ computation of the AND function with error
probability at most $\epsilon$ on every input.
Indeed, $k=\lceil\log(1/\epsilon)\rceil$ random parity checks on the
complemented input reduce the problem to the AND function on $k$ bits
with error at most
$2^{-k}$, and this smaller AND function can be computed exactly using
$2^k-1=O(1/\epsilon)$ qubits.

In the other direction, a periodic Fourier representation with $r$
nonconstant terms gives a torus approximation of degree logarithmic
in $r$ and in the inverse additional error. Here the representation
has fixed coefficients.

\begin{proposition}\label{prop:pfs-to-torus}
For every Boolean function $f\colon\{0,1\}^n\to\{0,1\}$ and
$\epsilon,\eta>0$ with $\sqrt{\epsilon}/2+\eta<1/4$,
\[
 \mathsf{deg}_{\mathbb{T},\sqrt{\epsilon}/2+\eta}(f/2)
 \le\log\frac{\max\{1,\pfs(f)\}}{2\eta}.
\]
\end{proposition}
\begin{proof}
Let $r=\pfs(f)$.
If $r=0$, a constant phase has error at most $\epsilon$, and
Fact~\ref{fact:torus-circle} gives a constant torus approximation
with error at most $\sqrt{\epsilon}/2$. Thus assume $r>0$.
Choose a periodic Fourier representation of minimum sparsity and write
\[
 P(u)=\alpha+\sum_{S\in\mathcal S}\beta_S\prod_{i\in S}u_i,
 \qquad
 \mathcal S\subseteq2^{[n]}\setminus\{\varnothing\},
 \qquad |\mathcal S|=r.
\]
Then $Q(x)=P(\phi^{\otimes n}(x))$ satisfies
$\operatorname{Err}(f(x)/2,Q(x))\le\epsilon$ on every input, and
\[
 Q(x)-\alpha=\sum_{S\in\mathcal S}\beta_S
       (-1)^{\sum_{i\in S}x_i}.
\]
Thus $Q-\alpha$ has Fourier sparsity at most $r$.
By Theorem~\ref{thm:f0}, there is a torus polynomial $R$ of degree
at most $\log(r/(2\eta))$ such that
\[
 \|Q(x)-\alpha-R(x)\|_{\mathbb{T}}\le\eta
 \qquad\forall x\in\{0,1\}^n.
\]
Set $\widetilde Q=\alpha+R$. Adding a constant does not increase
the degree, and for every $x\in\{0,1\}^n$,
\begin{align*}
 \|f(x)/2-\widetilde Q(x)\|_{\mathbb{T}}
 &\le\|f(x)/2-Q(x)\|_{\mathbb{T}}
       +\|Q(x)-\widetilde Q(x)\|_{\mathbb{T}}\\
 &\le\frac12\sqrt{\operatorname{Err}(f(x)/2,Q(x))}+\eta\\
 &\le\sqrt{\epsilon}/2+\eta.
\end{align*}
\end{proof}

\section{Conclusion}
\label{sec:conclusion}
We proved upper bounds on torus approximation degree from the Fourier
$\ell_1$ norm and from the size and depth of modular circuits.
The Fourier bound gives degree $O(\log(n/\epsilon))$ for
$\AND_n/2$, matching the known lower bound up to constant factors
at inverse-polynomial error and refuting the superlogarithmic-degree
conjectures of Krishan and Sarma.
For $\ACm$ circuits of size $s$ and depth $d$, the exponent of
$\log s$ is $\max\{1,\mu-1\}d$, where $\mu$ is the number of
distinct prime factors of $m$, and the bound is linear in
$\log(1/\epsilon)$.

Both arguments combine sampling with low-degree torus approximations
of weighted modular zero tests. In the Fourier argument, sampling
reduces the number of characters, and coefficient rounding allows
each weighted character to be represented with low degree modulo one.
For circuits, the modular approximation lemmas support an induction
on depth. A decomposition that alternates the order of the
prime-power factors gives the refined circuit exponent.

Periodic Fourier representations translate these approximation
bounds into quantum constructions. Functions in $\ACC$ admit
$\NMQC$ computations using quasipolynomially many qubits at
inverse-polynomial error, while Fourier sampling gives an
$O(n/\epsilon)$-qubit construction for the AND function.
These computations use fixed input parities and fixed measurements,
with error probability at most $\epsilon$ on every input.
Conversely, a periodic Fourier representation of sparsity $r\ge1$ with
error at most $\epsilon<1/16$ can be converted into a torus
approximation of degree at most $\log(r/\sqrt{\epsilon})$ with error
$\sqrt{\epsilon}$.
Hence, at inverse-polynomial error, polylogarithmic torus degree is
equivalent to quasipolynomial periodic Fourier sparsity.

\section*{Acknowledgments}
The work of RM was supported by JST FOREST Program Grant Number JPMJFR216V and
JSPS KAKENHI Grant Numbers JP22H00522 and JP26H02504.

\section*{AI disclosure}
ChatGPT (OpenAI) was used to assist with developing and checking
mathematical arguments, identifying relevant literature, and drafting
and revising the exposition throughout the paper, including the proofs
in the main text and appendices.
The authors take full responsibility for the correctness, originality,
and integrity of the paper, including its proofs and references.

\bibliographystyle{alpha}
\bibliography{biblio}

\clearpage
\appendix
\section{Refining the circuit exponent}
\label{app:refined-circuit-bound}
\subsection{A decomposition with fewer maps}
For $\mu=1$, Theorem~\ref{thm:intro} follows from
Proposition~\ref{prop:basic-circuit-bound}.
We prove the remaining case $\mu\ge2$ by decomposing the circuit into
Boolean maps associated with prime-power moduli and approximating
their composition.
The proof has two steps: a decomposition with at most
$(\mu-1)d+1$ maps, followed by a degree bound for substitution of
one map. The final map has a direct polynomial representation modulo
its modulus, so only the other maps contribute factors of $\log s$.
Write
\[
 m=\prod_{i=1}^{\mu}p_i^{e_i},
 \qquad q_i=p_i^{e_i}.
\]

The decomposition adapts the alternating prime order of Beigel and
Tarui's conversion to clean form
\cite[Section~5.2, p.~789]{beigel_tarui_1991}.
We keep the intermediate gate values as Boolean coordinates, so the
number of coordinates stays bounded by the original circuit size.

\begin{lemma}[Circuit decomposition with alternating prime-power order]\label{lem:circuit-decomposition}
Let $m$ have $\mu\ge2$ distinct prime factors.
There is a constant $t_m\ge1$ depending only on $m$ with the following
property. Every $n$-input $\ACm$ circuit $C$ of depth at most $d\ge1$
and size at most $s$ admits a factorization
\[
 C=B_H\circ\cdots\circ B_1,
 \qquad H\le(\mu-1)d+1,
\]
where $B_j\colon\{0,1\}^{r_{j-1}}\to\{0,1\}^{r_j}$,
$r_0=n$, $r_H=1$, and $r_j\le s$.
There is a map $\sigma\colon[H]\to[\mu]$ such that every coordinate
of $B_j$ has the form
\[
 (B_j)_a(y)=\prod_{\ell}\delta_{q_{\sigma(j)}}(G_{j,a,\ell}(y)),
\]
where $G_{j,a,\ell}\in\mathbb{Z}[y]$ and
$\deg G_{j,a,\ell}\le t_m$.
Moreover, there is an integer polynomial $Q$ of degree at most $t_m$
such that $B_H(y)\equiv Q(y)\pmod{q_{\sigma(H)}}$ on the Boolean cube.
\end{lemma}
\begin{proof}
If $C$ has no non-$\mathsf{NOT}$ gates, its output is a constant or
an input literal, and one map with $\sigma(1)=1$ suffices. Otherwise,
discard gates that do not feed the output and group the remaining gates by height,
where the height of a gate is the maximum number of non-$\mathsf{NOT}$
gates on a directed path ending at that gate, counting the gate itself.
Let $h\le d$ be the number of nonempty heights. Negations are incorporated into input literals
and, if necessary, into the final output.

Put
\[
 u_m=1+\max_{i\in[\mu]}e_i(q_i-1).
\]
For each $i\in[\mu]$ and any list of Boolean literals
$\ell_1(y),\ldots,\ell_u(y)$,
Lemma~\ref{lem:prime-power-representation}, with $k=e_i$, gives on
Boolean inputs
\[
 q_i Q_{q_i,e_i}\left(\sum_{a=1}^u\ell_a(y)\right)
 \equiv\delta_{q_i}\left(\sum_{a=1}^u\ell_a(y)\right)
 \pmod{q_i}.
\]
The polynomial on the left has degree at most $e_i(q_i-1)$
and takes integer values on Boolean inputs.
Its multilinearization therefore has integer coefficients: each
coefficient is an integer alternating sum of values at Boolean
vertices. Denote this integer polynomial by $R_i$.
Thus $R_i$ has degree at most $u_m-1$ and represents the
$\Mod{q_i}$ gate exactly modulo $q_i$.
In particular, its residues on Boolean inputs are zero or one.
The Chinese remainder theorem gives
\[
 \Mod m(\ell_1(y),\ldots,\ell_u(y))
 =\prod_{i=1}^{\mu}\delta_{q_i}
       \left(\sum_{a=1}^u\ell_a(y)\right).
\]

Process one height at a time, carrying all input bits and previously
computed gate values as coordinates. Replace a height by $\mu$ maps,
one for each prime-power factor in a chosen order.
For each new $\Mod m$ gate, maintain a Boolean coordinate $b$ for
the product of the tests processed so far.
At a stage with modulus $q_i$, set $b=R_i(y)\bmod q_i$ if it is
the first stage, and otherwise update
\[
 b\longmapsto bR_i(y)\bmod q_i.
\]
Here reduction modulo $q_i$ returns the Boolean residue zero or
one on every Boolean input to the map.
The update has an integer polynomial representation modulo $q_i$
of degree at most $u_m$.
Compute every AND gate of the height in the first map, using
\[
 \AND(\ell_1(y),\ldots,\ell_u(y))
 =\prod_{a=1}^u\delta_{q_i}(\ell_a(y)-1),
\]
and copy its value through the remaining maps. A partial product
uses the coordinate reserved for its original gate. Hence at most
$s$ coordinates are needed at any stage.

Every map has coordinates that are products of zero tests of
integer polynomials of degree at most $u_m$.
Indeed, a Boolean coordinate represented by $R$ modulo $q$ equals
$\delta_q(R-1)$. Copies are included by taking $R=y_a$.
Every map except the first at each height also has integer polynomial
coordinates modulo its modulus, of degree at most $u_m$, since it
only updates partial products and copies other coordinates.

Use the order
$(q_1,\ldots,q_\mu)$ at odd heights and the reverse order at
even heights. The last map at one height and the first map at the
next height use the same modulus $q$.
The former has integer polynomial coordinates modulo $q$.
Substituting those polynomials into the latter's zero tests gives
their exact composition, since polynomial substitution preserves
congruence modulo $q$. Each resulting test has degree at most $u_m^2$.
The $h-1$ pairs being merged are disjoint: each height has at least
two maps. Consequently, these substitutions do not accumulate, and
the total number of maps is
\[
 H=\mu h-(h-1)=(\mu-1)h+1\le(\mu-1)d+1.
\]
The final map is unmerged and has polynomial coordinates modulo its
modulus. Select the circuit output coordinate there, applying
$1-y_a$ if the output is negated. This gives the required scalar
polynomial $Q$.

For each resulting map $B_j$, define $\sigma(j)=i$ when its
modulus is $q_i$. Taking $t_m=u_m^2$ completes the proof.
\end{proof}

\subsection{Substituting a Boolean map}
We use the character expansion of products of modular zero
tests~\cite{chen_papakonstantinou_2019} to approximate $P\circ B$.
As in Section~\ref{sec:acc-torus}, we sample monomials and characters
jointly. The resulting degree depends logarithmically on the number
of monomials, without a factor for the number of zero tests in each
coordinate.

\begin{lemma}[Substitution of a Boolean map]\label{lem:block-substitution}
Let $q=p^e$ be a prime power, let $n,r,t\ge1$, and let
$B\colon\{0,1\}^n\to\{0,1\}^r$ have coordinates
\[
 B_a(x)=\prod_j\delta_q(G_{a,j}(x)),
 \qquad G_{a,j}\in\mathbb{Z}[x],\quad \deg G_{a,j}\le t.
\]
For every torus polynomial $P$ of degree at most $D\ge1$ in $r$
variables and every $\eta\in(0,1/2)$, there is a torus polynomial $R$
such that
\[
 \|R(x)-P(B(x))\|_{\mathbb{T}}\le\eta
 \qquad\forall x\in\{0,1\}^n
\]
and
\[
 \deg R\le c_q t\bigl(D\log(r+1)+\log(n+1)+\log(1/\eta)\bigr),
\]
where $c_q$ depends only on $q$.
\end{lemma}
\begin{proof}
Reduce the coefficients of a multilinear representative of $P$
modulo one and write
\[
 P(y)=\sum_{S\in\mathcal U}\alpha_S y_S\pmod1,
 \qquad y_S=\prod_{a\in S}y_a,
 \qquad 0<|\alpha_S|\le1/2.
\]
Here $|S|\le D$ for every $S\in\mathcal U$.
Set $K=|\mathcal U|$ and $A=\sum_S|\alpha_S|$. Then
$K\le(r+1)^D$ and $A\le K/2$.
If $A=0$, take $R=0$. Otherwise, define the real-valued function
\[
 F(x)=\sum_{S\in\mathcal U}\alpha_S b_S(x),
 \qquad b_S(x)=\prod_{a\in S}B_a(x).
\]
It represents $P(B(x))$ modulo one. Every $b_S$ is a product of
zero tests $\delta_q(G_{S,j})$ with $\deg G_{S,j}\le t$.
Character orthogonality gives
\[
 b_S(x)=\E_{\lambda\in\{0,\ldots,q-1\}^{u_S}}
 \left[\cos\left(\frac{2\pi}{q}
                  \sum_{j=1}^{u_S}\lambda_jG_{S,j}(x)\right)\right].
\]

To approximate $F$ by a short sum, we combine these expectations into
one and then approximate it by a sample average.
Sample an index $S\in\mathcal U$ with probability $|\alpha_S|/A$,
and conditional on $S$, sample $\lambda$ uniformly from
$\{0,\ldots,q-1\}^{u_S}$.
Define the random function
\[
 Z_{S,\lambda}(x)=\sgn(\alpha_S)
 \cos\left(\frac{2\pi}{q}\sum_j\lambda_jG_{S,j}(x)\right).
\]
Then $|Z_{S,\lambda}(x)|\le1$ and
\[
 A\E[Z_{S,\lambda}(x)]
 =A\sum_{S\in\mathcal U}\frac{|\alpha_S|}{A}
       \sgn(\alpha_S)b_S(x)
 =F(x).
\]
Take $M=\lceil8(n+1)A^2/\eta^2\rceil$ independent samples.
For each fixed input, Hoeffding's inequality~\cite{hoeffding_1963}
gives
\[
 \Pr\left[\left|\frac A M\sum_{j=1}^M Z_j(x)-F(x)\right|>\eta/2\right]
 \le2\exp\left(-\frac{M\eta^2}{8A^2}\right)
 \le2\exp(-(n+1))<2^{-n}.
\]
A union bound over all $2^n$ inputs therefore lets us fix samples
such that their weighted average $\widetilde F$ satisfies
$|\widetilde F(x)-F(x)|\le\eta/2$ everywhere.

Each sample is a signed cosine of an integer polynomial $G$ of
degree at most $t$. Using
\[
 \cos(2\pi G(x)/q)
 =\sum_{a=0}^{q-1}\cos(2\pi a/q)\delta_q(G(x)-a),
\]
write $\widetilde F$ as a sum of at most $Mq$ weighted zero tests.
Apply Lemma~\ref{lem:apxpe} to each test with error $\eta/(2Mq)$
and sum the resulting polynomials. Their sum $R$ approximates
$\widetilde F$ with torus error at most $\eta/2$, so it approximates
$P\circ B$ with error at most $\eta$.
Its degree is at most
\[
 t(q-1)\log(2Mq/\eta).
\]
Since $M\le3(n+1)K^2/\eta^2$, this is at most
\[
 t(q-1)\bigl(2D\log(r+1)+\log(n+1)
                 +3\log(1/\eta)+\log(6q)\bigr).
\]
Absorbing constants into $c_q$ proves the claim.
\end{proof}

\subsection{Proof of Theorem~\ref{thm:intro}}
\begin{circuitrestatement}[Restated]
For every $m\ge2$, there is a constant $C_m\ge1$ depending only on
$m$ such that the following holds. Let $\mu$ be the number of distinct
prime factors of $m$ and set $\nu=\max\{1,\mu-1\}$.
For every $n\ge1$, every $n$-input $\ACm$ circuit $C$ of depth at most $d$ and size
at most $s\ge2$, and every $\epsilon\in(0,1/4)$,
\[
 \tdeg(C/2)\le C_m^d(\log s)^{\nu d}\log(1/\epsilon).
\]
\end{circuitrestatement}
\begin{proof}
If $\mu=1$, apply Proposition~\ref{prop:basic-circuit-bound}
with $C_m=A_m$. Assume henceforth that $\mu\ge2$, so $\nu=\mu-1$.
Set $L=\log s\ge1$.
If $C$ has no non-$\mathsf{NOT}$ gates, $C/2$ has an exact affine
representation, and the bound follows. This includes $d=0$.
Otherwise, apply Lemma~\ref{lem:circuit-decomposition} to write
$C=B_H\circ\cdots\circ B_1$ with $H\le(\mu-1)d+1$.
For $1\le j\le H$, let
\[
 F_j=B_H\circ\cdots\circ B_{H-j+1}.
\]
We prove that there is a constant $a_m\ge1$ such that, for every
$\eta\in(0,1/4)$, $F_j/2$ has an $\eta$-approximating torus polynomial
of degree at most
\begin{equation}\label{eq:block-degree}
 a_m^jL^{j-1}\log(1/\eta).
\end{equation}

For $j=1$, the final map has a polynomial representation $Q$ modulo
$q_{\sigma(H)}$ of degree at most $t_m$.
Since $F_1$ is Boolean and $Q\equiv F_1\pmod{q_{\sigma(H)}}$,
\[
 F_1=\delta_{q_{\sigma(H)}}(Q-1).
\]
Lemma~\ref{lem:apxpe}, with weight $1/2$, gives degree at most
$t_m(q_{\sigma(H)}-1)\log(1/\eta)$, proving the base case for a
sufficiently large $a_m$.

Suppose the claim holds for $j<H$ and put $E_\eta=\log(1/\eta)$.
Choose an $(\eta/2)$-approximation $P$ to $F_j/2$ by the induction
hypothesis and put $D=\max\{1,\deg P\}$. Then
\[
 D\le a_m^jL^{j-1}\log(2/\eta)
 \le2a_m^jL^{j-1}E_\eta.
\]
Apply Lemma~\ref{lem:block-substitution} to
$P\circ B_{H-j}$ with modulus $q_{\sigma(H-j)}$ and error $\eta/2$.
All input and output dimensions are at most $s$, all test degrees
are at most $t_m$, and all moduli divide $m$.
Since $\log(s+1)\le2L$, the resulting polynomial has degree at most
\[
 b_m\bigl((D+1)L+\log(2/\eta)\bigr)
 \le5b_m a_m^jL^jE_\eta
\]
for a constant $b_m$ depending only on $m$.
Its error relative to $F_{j+1}/2$ is at most $\eta$, by the triangle
inequality. Taking $a_m\ge5b_m$ proves \eqref{eq:block-degree} for
$j+1$.

Finally, $H-1\le(\mu-1)d$ and $H\le\mu d$, since $d\ge1$.
Applying \eqref{eq:block-degree} with $j=H$ and $\eta=\epsilon$ gives
\[
 \tdeg(C/2)\le a_m^HL^{H-1}\log(1/\epsilon)
 \le C_m^dL^{(\mu-1)d}\log(1/\epsilon),
\]
where $C_m=a_m^\mu$.
\end{proof}

\section{Functional degree and polynomial representations}
\label{app:functional-degree}
The upper bound $(q-1)k$ in
Lemma~\ref{lem:prime-power-representation} is sufficient for the main
proofs. Here we relate functional degree to polynomial representation
and record the exact degree for prime-power zero tests.

Aichinger and Moosbauer~\cite{aichinger_moosbauer_2021} develop a
finite-difference calculus for maps between abelian groups.
Clark and Schauz~\cite{CLARK2022691} determine the possible functional
degrees of such maps, including the exact degree used below.
For related results on polynomial representations modulo prime
powers, see Wilson~\cite{wilson_2006}.

For abelian groups $A,B$ and a map $f\colon A\to B$, write
$\Delta_hf(z)=f(z+h)-f(z)$.
For nonzero $f$, its functional degree $\fdeg(f)$ is the least
$d\ge0$ such that $\Delta_{h_1}\cdots\Delta_{h_{d+1}}f=0$ for every
$h_1,\ldots,h_{d+1}\in A$, or $+\infty$ if no such $d$ exists.
We set $\fdeg(0)=-\infty$.

When $A=\mathbb{F}_p^r$, $B=\mathbb{T}$, and $\fdeg(f)<+\infty$,
the functional degree of $f$ agrees with its degree as a
non-classical polynomial~\cite[Definition~1.2]{tao_inverse_2012}.
Classical $\mathbb{F}_p$-valued polynomials are included through the
embedding $a\mapsto a/p\pmod1$.
Below we apply the same finite-difference definition to the cyclic
group $\mathbb{Z}/q\mathbb{Z}$ and relate it to ordinary univariate
polynomial degree.

\begin{proposition}[Polynomial representation]\label{prop:polynomial-representation}
Let $q\ge1$ and $d\ge0$ be integers, and let
$f\colon\mathbb{Z}/q\mathbb{Z}\to\mathbb{T}$.
Then $\fdeg(f)\le d$ if and only if there is a real polynomial
$P\in\mathbb{R}[z]$ of degree at most $d$ such that
\[
 P(z)=f(z\bmod q)\pmod1
 \qquad\forall z\in\mathbb{Z}.
\]
\end{proposition}
\begin{proof}
Let $F(z)=f(z\bmod q)$ and $\Delta F(z)=F(z+1)-F(z)$.
If $\fdeg(f)\le d$, then $\Delta^{d+1}F=0$.
The finite Newton expansion, valid also for negative integers, gives
\[
 F(z)=\sum_{j=0}^d\binom zj(\Delta^jF)(0)
 \qquad\forall z\in\mathbb{Z}.
\]
Indeed, on the subgroup generated by the translates of $F$, the
shift $T=I+\Delta$ satisfies
$T^z=\sum_{j=0}^d\binom zj\Delta^j$ for every integer $z$.
Choosing real representatives $c_j$ of $(\Delta^jF)(0)\in\mathbb{T}$
therefore gives $P(z)=\sum_{j=0}^d c_j\binom zj$ as required.
Conversely, each additive derivative lowers the degree of a real
polynomial by at least one, so any $d+1$ additive derivatives of
$f$ vanish if such a $P$ exists.
\end{proof}

\begin{proposition}[{\cite[Proposition~4.4 and Theorem~4.7]{CLARK2022691}}]
\label{prop:delta-functional-degree}
For every prime $p$ and integers $e,k\ge1$,
\[
 \fdeg\left(\frac{\delta_{p^e}}{p^k}\right)
 =(p-1)p^{e-1}k+p^{e-1}-1.
\]
Here the domain is $\mathbb{Z}/p^e\mathbb{Z}$ and the codomain is
$\mathbb{T}$.
\end{proposition}
\begin{proof}
Clark and Schauz give this degree for the delta function
$\mathbb{Z}/p^e\mathbb{Z}\to\mathbb{Z}/p^k\mathbb{Z}$.
The injective homomorphism $a\mapsto a/p^k\pmod1$ from
$\mathbb{Z}/p^k\mathbb{Z}$ into $\mathbb{T}$ commutes with additive
derivatives and therefore preserves functional degree.
\end{proof}

Put $q=p^e$. By Proposition~\ref{prop:polynomial-representation}, this functional
degree is also the minimum degree of a real polynomial representing
$\delta_q/p^k$ modulo one on the integers.
If this functional degree is $D$, then
$\Delta^j(\delta_q/p^k)=0$ in $\mathbb T$ for every $j>D$.
In particular, $(\Delta^j\delta_q)(0)\equiv0\pmod{p^k}$ for every
$j>D$.
The coefficients $a_j$ chosen in
Lemma~\ref{lem:prime-power-representation} are then zero, so the
polynomial $Q_{q,k}$ constructed there has degree at most this value.
Minimality gives the reverse inequality. Thus that polynomial
attains the optimal degree
\[
 \deg Q_{q,k}=(p-1)p^{e-1}k+p^{e-1}-1
 =(q-1)k-(p^{e-1}-1)(k-1)\le(q-1)k,
\]
with strict inequality when $e,k>1$.
For $e=1$, the exact degree is $k(p-1)$, which is also the
non-classical degree of $\delta_p/p^k$ on $\mathbb{F}_p$.
The approximation arguments in the main text use the simpler bound
$(q-1)k$.

\end{document}